# Detecting orbital angular momentum through division-of-amplitude interference with a circular plasmonic lens

Ai-Ping Liu,[1] Xiao Xiong,[1] Xi-Feng Ren,[1, *] Yong-Jing Cai,[1] Guang-Hao Rui,[2] Qi-Wen Zhan,[3, #] Guang-Can Guo,[1] and Guo-Ping Guo[1]

[1]Key Laboratory of Quantum Information, University of Science and Technology of China, Hefei, Anhui 230026, China, [2]Department of Optics and Optical engineering, University of Science and Technology of China, Heifei, Anhui 230026, China, [3]Electro-Optics Program, University of Dayton, 300 College Park, Dayton, Ohio 45469, USA

*renxf@ustc.edu.cn，#qzhan1@udayton.edu

**We demonstrate a novel detection scheme for the orbital angular momentum (OAM) of light using circular plasmonic lens. Owing to a division-of-amplitude interference phenomenon between the surface plasmon waves and directly transmitted light, specific intensity distributions are formed near the plasmonic lens surface under different OAM excitations. Due to different phase behaviors of the evanescent surface plasmon wave and the direct transmission, interference patterns rotate as the observation plane moves away from the lens surface. The rotation direction is a direct measure of the sign of OAM, while the amount of rotation is linked to the absolute value of the OAM. This OAM detection scheme is validated experimentally and numerically. Analytical expressions are derived to provide insights and explanations of this detection scheme. This work forms the basis for the realization of a compact and integrated OAM detection architect that may significantly benefit optical information processing with OAM states.**

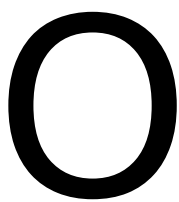Orbital angular momentum is an intrinsic property of light that has attracted attentions in many recent studies[1]. Benefiting from the specific doughnut transverse intensity profiles and helical wavefronts as shown in Fig. 1 (a), light waves carrying OAMs have been used in many areas, such as optical trapping, imaging, optical communication and even quantum information processing [2-10]. Efficient and robust detection of OAM states carried by

the vortex beam is therefore meaningful but remains challenging. Several schemes that based on Mach-Zehnder interferometer, deflection of a quartz wave plate, refractive or holographic elements[11-14], have been proposed to detect the OAM. However, these methods involve the use of bulky elements and complicated procedures that are hard to be integrated into a compact platform. Recently, plasmonic photodiode using sub-wavelength holographic metallic structures for OAM detection is reported in Ref. 15. The approach enables the realization of a compact and integrated OAM detection scheme, but only works for one specific OAM state. In this work, we propose a simple yet efficient OAM detection scheme based on a near-field division-of-amplitude interference phenomenon arising from a plasmonic lens which is made of a single circular groove etched into gold film. Different OAM states can be successfully distinguished by imaging the intensity distributions near the plasmonic lens surface.

Surface plasmon polaritons (SPPs) are the collective oscillations of free electrons and propagate along the interface between metal and dielectric[16]. In the past decade, the research of SPPs has attracted significant interests due to their fascinating properties such as high sensitivity to ambient refractive index and high local field enhancement etc. Diverse applications based on their unique characteristics, such as enhanced transmission, Raman and fluorescence enhancement, logic gate, information encoding, etc., have been proposed and demonstrated[17-25]. The relationship and interactions between SPPs and optical OAM states also attracted increasing attentions recently. Studies have shown that SPPs could potentially play an important role in the applications of OAM for information processing[25, 26]. When excited by vortex beam carrying OAM, the helical phase information can be imparted onto the excited SPPs and plays a decisive role of the interference profile near the focus of the plasmonic lens[27-32]. When light incidents on a plasmonic lens that is made of open slots etched into a metallic thin film, part of the incident light couples to the SPPs that propagate in the plasmonic lens surface and evanescently decays away from the surface, while part of incident light directly transmits through the slot and continues to propagate in the free space[33, 34]. Near the plasmonic lens surface, both the evanescent surface SPP wave and the directly transmitted propagating wave form their own field patterns. These two patterns will interfere with each other and create a unique intensity profile that is highly dependent on the OAM state of the incident light. More interestingly, as the observation plane moves away from the surface of the plasmonic lens, the field arising from the SPPs decays exponentially in amplitude but maintains a constant phase pattern, while the directly transmitted field will accumulate phase shift due to the propagation. Such a difference in the phase behavior of the two contributing components causes a rotation of the interference pattern that also depends on the OAM state of the incident light. The characteristics of this division-of-amplitude interference phenomenon arising from the plasmonic lens form the foundation of the OAM detection scheme we propose in this work.

In this paper, we first show the experimental observations of the intensity patterns near the surface of a plasmonic lens under the illumination of linearly polarized

Laguerre-Gaussian (LG) modes carrying specific OAM states. Their corresponding intensity profiles formed by the interference between the evanescent surface wave and the directly transmitted light are obtained by a collection mode near-field scanning optical microscope (NSOM). Analytical expressions of the two constituting field components are then derived to provide insights and physical explanations of the experimentally observed patterns. Numerical calculations are also performed on the platform of COMSOL Multiphysics 4.2 as validation of the analytical predictions and comparison with the experimental results. These results clearly demonstrate that different OAMs can be distinguished by analyzing the intensity distributions near the surface of the plasmonic lens, enabling an efficient and robust method for OAM detection.

## Results

**Experiment setup and results.** The vortex beam carrying different OAMs ($l\hbar$ per photon) has different wavefront phase variations as shown in Fig. 1(a) and transverse intensity profiles (shown in the insets of Fig. 1(a)). The sample used in the experiment is prepared on a 135-nm gold thin film evaporated on a $SiO_2$ substrate as shown in Fig. 1(b). A ring-shaped slot with a 200 nm width and a 70 nm depth is etched into the gold film by focusing ion beam (FIB) lithography, forming a plasmonic lens with an inner diameter of 6 μm. The inset 2 of Fig. 1(b) is the scanning electronic microscope (SEM) image of the plasmonic lens. A NSOM probe working on collection mode scans over the plasmonic lens with a constant tip-sample distance $h = 200$ nm. Then the Multiviews

4000™ system forms the NSOM images of the intensity distribution of the electric field above the plasmonic lens.

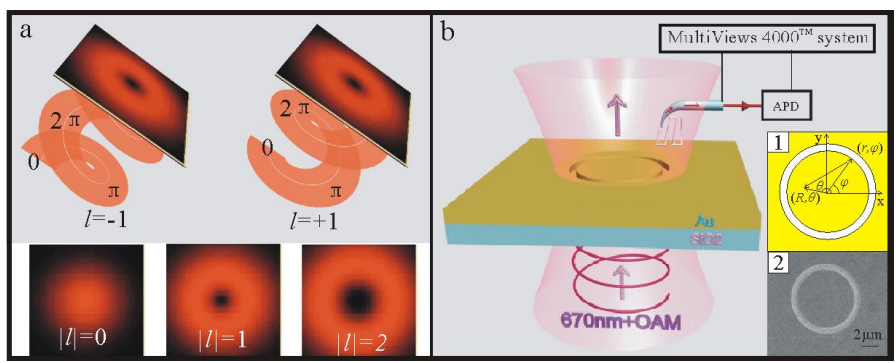

Figure 1. (a). Schematic of the LG modes. The wavefront phase variation is different for LG modes carrying different OAMs. The insets are the transverse intensity profiles of LG modes with different |l|. (b). Schematic of experimental setup. The plasmonic lens is excited by LG mode from the $SiO_2$ substrate side and imaged by a NSOM probe working on collection mode. Inset 1 is the diagram of a single ring plasmonic lens and the coordinates used in analytical derivation. The illumination is along the z-direction. Inset 2 is the SEM micrograph of the plasmonic lens fabricated in gold film on $SiO_2$ substrate.

The intensity distributions near the plasmonic lens surface excited by photons with different OAMs are summarized in Fig. 2 (a)-(e) with the corresponding COMSOL numerical modeling results shown in Fig. 2(f)-(j) for comparison. The polarization direction of the incident light is indicated by the white arrow shown in Fig. 2 (a). It should be noted that the specific NSOM probe we use, which is a $SiO_2$ probe coated with a 35-nm Au/Cr layer and a tip diameter of 100 nm, mainly collects the transversal electric field[36, 37]. Thus only the intensity distributions of the transverse component are shown in the numerical simulation results as well. Very good agreements are obtained between the experimental and numerical results. For all of the intensity distributions, one common prominent feature is the bright arc that converges towards the center. This is due to the fact that only the component of the linearly polarized incident that is TM polarized locally with respect to the circular slot can be coupled to the SPPs. The spacing between adjacent fringes of the arc pattern is 320 nm, which corresponds to half of the SPP wavelength. Closer examination of the intensity patterns also reveals weaker circular patterns that converge towards the center, which can be attributed to the directly transmitted light.

Clear differences are observed from these intensity distributions under illumination with different OAM states. For $l = 0$, which corresponds to a fundamental Gaussian illumination, the transverse components of the SPP field from the opposite sides of the circular plasmonic lens interfere constructively in the center and form a solid spot. For $l \neq 0$, the intensity distributions differ not only along the radial direction but also along the azimuthal direction. A dark spot appears for $|l| = 1$ due to destructive interferences while a bright spot shows up again for $|l| = 2$, as shown in Fig. 2. What is more interesting is the rotating intensity distribution in the center with the rotation direction and amount directly related to OAM. The numerical calculation results (Fig. 2(f)-(j)) confirm this fact. For $l = +1$, there are two separated semi-arcs in the center which rotate anti-clockwise. Once the OAM changes to $l = -1$, the rotation direction becomes clockwise with the same angle. In the cases of $l = \pm 2$, there is an elliptical solid spot in the center, which is different from the cases of $l = \pm 1$. But the rotation of the intensity distribution in the center is similar to the cases of $|l| = 1$, with anti-clockwise rotation for $l > 0$ and clockwise rotation for $l < 0$.

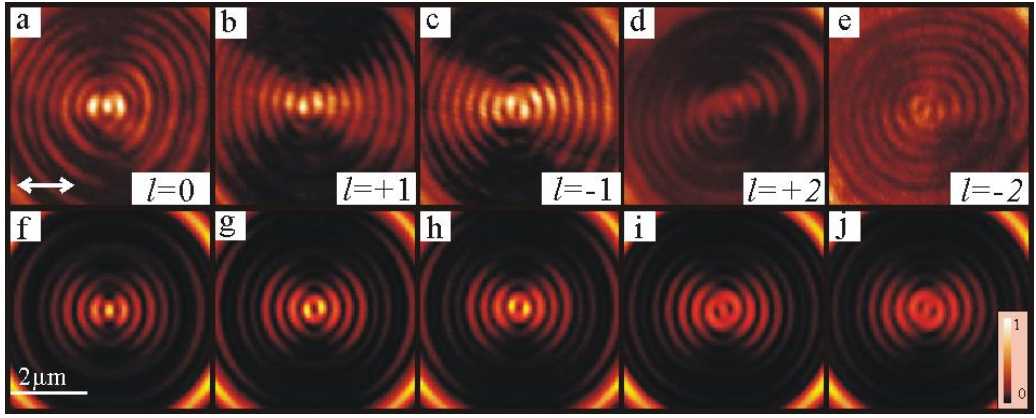

Figure 2. Intensity distributions of the optical field near the plasmonic lens surface excited by photons with different OAMs ($h = 200$ nm). (a)-(e) are the NSOM images for $l = 0, +1, -1, +2$ and $-2$, respectively. (f)-(j) are the corresponding numerical simulation results. The excitation polarization is shown by the white arrow in (a). The scale bar in (f) and the color bar in (j) are also applied for the other images of intensity distributions.

**Explanations and discussions.** The above phenomenon can be understood by the interference between the evanescent SPP field and the directly transmitted light arising from the circular plasmonic lens. The coordinates for the analytical derivation are illustrated in the inset (2) of Fig. 1(b). A single ring slot is etched into a thin metal film deposited on glass substrate. Light illuminates the structure normally from the substrate side. Considering an incident light linearly polarized in the x-direction with OAM of $l\hbar$ ($l = 0, \pm 1, \pm 2 \cdots$), it can be expressed as:

$$E_{inc} = exp(il\varphi) \cdot \hat{e}_x, \qquad (1)$$

where $exp(il\varphi)$ is the phase due to the OAM. The subwavelength slot opening can be regarded as an array of secondary sources. For a sufficiently narrow slot, only the radial component can couple to excite SPPs. Thus the z-component of plasmonic field at an observation point $(R, \theta)$ near the origin due to the excitation along an incremental length of the annular slot is given by

$$dE_{spp,z} = \hat{e}_z E_{0z} cos\varphi \cdot exp(-k_{spp,z}z) \cdot exp[il\varphi + ik_{spp,r}(R\hat{e}_R - r\hat{e}_r)] rd\varphi, \quad (2)$$

where $E_{0z}$ is a constant that is related to the coupling efficiency, $r$ is the radius of the nano-ring and $k_{spp,r}$ is the wave vector of the SPPs that propagate from the excitation location to the center, $k_{spp,z}$ is the wavenumber of the SPPs that describes its decaying amplitude in the z-direction. The factor of $cos\varphi$ is due to the fact that only the electric field that is locally TM polarized (the radial component) with respect to the slit can be coupled into the SPPs.

For each of these secondary SPP waves, there is also a transverse component lying in the propagation direction towards the center of the structure:

$$dE_{spp,r} = \hat{e}_r \left(i\frac{k_{spp,z}}{k_{spp,r}}\right) E_{0z} cos\varphi\, exp(il\varphi) exp[il\varphi + ik_{spp,r}(R\hat{e}_R - r\hat{e}_r)] rd\varphi. \quad (3)$$

The x-polarized plasmonic field at the observation point can be expressed as

$$E_{spp,x}(R,\theta) = \hat{e}_x \left(i\frac{k_{spp,z}}{k_{spp,r}}\right) E_{0z} exp(-k_{spp,z}z) exp(ik_{spp,r}r) \cdot \int_0^{2\pi} cos^2\varphi\, exp[il\varphi + ik_{spp,r}Rcos(\theta - \varphi)] rd\varphi, \quad (4)$$

where another factor of $cos\varphi$ is due to the projection of the transverse field component to the x-direction. For simplicity, we neglect the propagation loss of the SPPs, i.e. $\text{Im}(k_r) = 0$, and then $k_{spp,r} = -k_{spp,r}\hat{e}_r = -2\pi/\lambda_{spp}\hat{e}_r$. With these assumptions, Eq. (4) can be simplified as:

$$E_{spp,x}(R,\theta)$$
$$= \hat{e}_x 2\pi i^l \left(i\frac{k_{spp,z}}{k_{spp,r}}\right) E_{0z} exp(-k_{spp,z}z) exp(ik_{spp,r}r)\left\{\frac{1}{2}exp(il\theta)J_l(ik_{spp,r}R)\right.$$
$$\left. -\frac{1}{4}exp[i(l+2)\theta]J_{l+2}(k_{spp,r}R) -\frac{1}{4}exp[i(l-2)\theta]J_{l-2}(k_{spp,r}R)\right\}, \qquad (5)$$

where $J_n$ is the nth order Bessel function of the first kind. This simple analytical derivation clearly indicates that the x-polarized component of the plasmonic field near the center of the nano-ring consists of three evanescent components with different topological charges of $l$, $l+2$ and $l-2$.

Besides the surface SPP waves, there is also finite amount of direct transmission of the incident light through the slit. This directed transmitted light will be substantially polarized along the same direction of the illumination (x-direction). Similar to the derivation above, the directly transmitted x-polarized field from an incremental length of the annular slot can be written as

$$dE_t = \hat{e}_x E_{0x} exp(ik_z z) exp[il\varphi + ik_r(R\hat{e}_r - r\hat{e}_r)] r d\varphi, \qquad (6)$$

where $E_{0x}$ is a constant that is related to the transmission efficiency, $r$ is the radius of the nano-ring and $k_r$ is the transverse wave vector of the propagating light wave towards the center, and $k_z$ is the z-component of the wave vector of the transmitted propagating light wave.

The x-polarized directly transmitted field at the observation point can be expressed as

$$E_t(R,\theta) = \hat{e}_x E_{0x} exp(ik_z z) exp(ik_r r) \int_0^{2\pi} exp[il\varphi + ik_r R cos(\theta - \varphi)] r d\varphi, \qquad (7)$$

$$E_t(R,\theta) = \hat{e}_x 2\pi i^l E_{0x} exp(ik_z z) exp(ik_r r) exp(il\theta) J_l(k_r R). \qquad (8)$$

Clearly, the directly transmitted field has only one component with topological charge of $l$. The directly transmitted field and the surface SPP field will interfere with each other. In general, the difference between the $k_{spp,r}$ and $k_r$ are very small at the metal/dielectric surface. For a small device, this difference can be ignored in terms. For $l = 0$, and $\pm 2$, a $J_0$ term will show up in Eq. (5), leaving a bright spot in the center. For $l = \pm 1$ and any other values, no $J_0$ will show up in Eqs. (5) and (8), thus the center of the resulted field will be always dark. This exactly explains the experimental and numerical simulation results shown in Fig. 2. As one moves away from the surface, the phase pattern of the SPP field does not change due to the evanescent feature. However, the directly transmitted component will accumulate phase changes as it propagates away from the surface. The phase shift of the directly transmitted component will cause the overall interference patterns to rotate. The sense of rotation depends on the sign of the topological charge of the illumination. The amount of rotation depends on distance away from the surface (the phase accumulated through propagation of the directly transmitted field), and the relative strengths between the surface SPP field and the directly transmitted light (the coupling efficiency, transmission efficiency and the ratio of $k_{spp,z}/k_{spp,r}$), which can be engineered by the nano-ring plasmonic lens design.

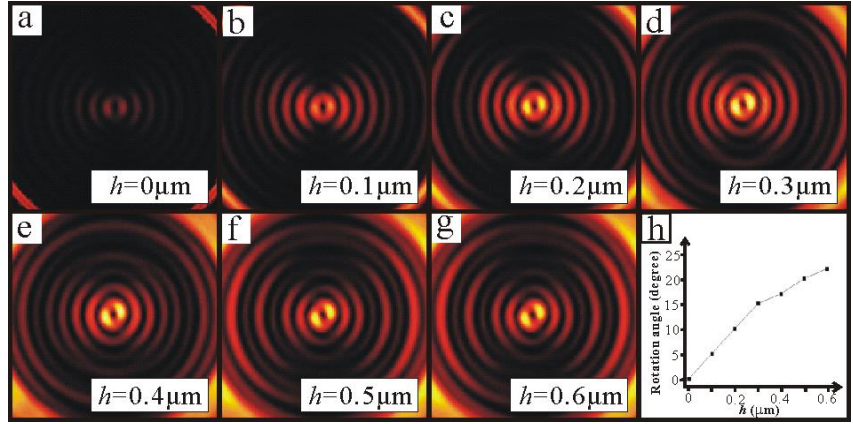

Figure 3. Rotation of the intensity distributions in the center with different tip-sample distance $h$ for $l = +1$. (a) - (g). $h = 0$ μm, 0.1 μm, 0.2 μm, 0.3 μm, 0.4 μm, 0.5 μm and 0.6 μm. (h). The rotation angle versus the tip-sample distance $h$. The excitation polarization is shown by the white arrow in Fig. 2(a).

To confirm the analysis given above, Fig. 3(a)-(g) illustrate the intensity distributions of the electric field for $l = +1$ with different distances from the plasmonic lens surface. The intensity distribution is axial symmetric for $h = 0$, which is just on the interface between gold and air. When the tip-sample distance $h$ increases, the rotation angle increases as shown in Fig. 3(h). It is worth noting that this rotation effect only occurs at the near field range, since the SPPs decay exponentially with $h$.

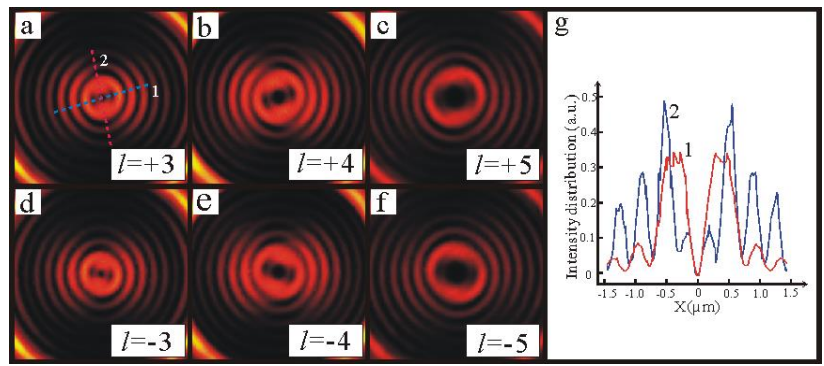

Figure 4. Numerical simulations of the intensity distributions with an elliptical vacant spot in the center for different OAMs ($h = 200$ nm). (a) $l = +3$, (b) $l = +4$, (c) $l = +5$, (d) $l = -3$, (e) $l = -4$, and (f) $l = -5$. The excitation polarization is shown by the white

arrow in Fig. 2(a). (g) Line scans of the intensity distribution along line 1 and line 2 in (a).

Table 1. The Diameters of the Dark Center Spots under Different OAM Excitations

| $|l|$ | 3 | 4 | 5 |
|---|---|---|---|
| $d_1$ (μm) | 0.15 | 0.36 | 0.59 |
| $d_2$ (μm) | 0.27 | 0.54 | 0.76 |

For higher OAM states, the center of the resulted interference pattern will be always dark. To illustrate thus, numerical simulation results for $|l| = 3, 4, 5$ with $h = 200$nm are summarized in Fig. 4. All have a dark valley in the center surrounded by an elliptical bright ring. However, the size of the dark core increases with the topological charge. In addition, the interference pattern still rotates as one moves away from the surface with the dark valley rotates anti-clockwise for $l = +3, +4, +5$, and clockwise for $l = -3, -4, -5$. Line scans of the intensity distributions along the major and minor axes (line 1 and line 2 indicated in Fig. 4(a)) for $l = +3$ are shown in Fig. 4(g). The full-width-at-half-maximums (FWHMs) of the dark center along the minor axis $d_1$ and major axis $d_2$ for $|l| = 3, 4, 5$ are summarized in Table 1. Clearly both $d_1$ and $d_2$ increase with the increase of OAM.

From the above analysis, we can conclude that the intensity distributions of the electromagnetic field near the plasmonic lens surface under different OAM excitations are different from each other, which can be exploited for efficient detection scheme of OAM states. The rotation direction of the intensity distribution is a direct measure of the sign of OAM, while the shape and amount of rotation are linked to the absolute value of the OAM. That is to say, for a specific incident OAM state, the electromagnetic field on the plasmonic lens has unique intensity distribution pattern. The detection of OAM can be done by analyzing the intensity distributions of the electric field near the plasmonic lens surface. This detection scheme may be further optimized through adjusting the relative strengths between the surface SPP field and the directly transmitted light by engineering the nano-ring plasmonic lens (ring size, slot depth and width, choice of materials etc.). It is worthy of pointing out that, although NSOM images are used for the experimental confirmation, it is conceivable that an integrated diode array can be used instead, enabling a much more miniaturized OAM detection architect.

**Discussion**

In summary, we presented a novel OAM detection scheme using the division-of-amplitude interference phenomenon between the surface plasmon waves and directly transmitted light arising from a circular plasmonic lens. Specific intensity distributions are formed near the plasmonic lens surface under different OAM excitations. The OAM state of the excitation light can be distinguished by analyzing the intensity distributions near the plasmonic lens. This OAM detection scheme is validated experimentally with NSOM images and numerical modeling results. Analytical expressions are derived to provide insights and explanations of this detection scheme. Very good agreements have been obtained among the experimental results, the analytical predictions and numerical simulations. The principle reported in this work enables the design of a compact and integrated OAM detection architect that may find important applications in optical communications and information processing with OAM states.

## Methods

The excitation LG mode used in the experiment is realized by transmitting a linearly polarized Gaussian beam through a computer generated hologram. The sample used in the experiment is prepared on a 135-nm gold thin film evaporated on a $SiO_2$ substrate. A ring-shaped slot with a 200 nm width and a 70 nm depth is etched into the gold film by FIB lithography, forming a plasmonic lens with an inner diameter of 6 μm.

A linearly polarized LG mode at 670 nm with different OAMs illuminates the plasmonic lens vertically from the $SiO_2$ substrate side using an objective lens ($10\times, N.A.= 0.25$). A NSOM probe working on collection mode scans over the plasmonic lens with a constant tip-sample distance $h = 200$ nm. The collected signal is guided to an Avalanche Photo Diode by a single mode polarization maintaining fiber. Then the Multiviews 4000™ system forms the NSOM images of the intensity distribution of the electric field above the plasmonic lens.

## Acknowledgements

This work was funded by the National Basic Research Program of China (Grants No. 2011CBA00200 and 2011CB921200), the Innovation funds from Chinese Academy of Sciences (Grants No. 60921091), the National Natural Science Foundation of China (Grants No. 10904137 and 10934006), and the Fundamental Research Funds for the Central Universities (Grants No.WK2470000005). X. F. Ren thanks the Program for New Century Excellent Talents in University.

## Author contributions

A.P.L., Y.J.C and X.F.R planned and performed the experiments, collected and analyzed the data, and wrote the paper. X.X., G.H.R, G.C.G and G.P.G. analyzed the data and gave the numerical calculations. Q.W.Z gave the theoretical explanations. X.F.R. and Q.W.Z. supervised the project, and conceived the experiments, analyzed the results and wrote the paper.

## Additional information

The authors declare no competing financial interests.